# STATUS OF HIGH TEMPERATURE SUPERCONDUCTOR BASED MAGNETS AND THE CONDUCTORS THEY DEPEND UPON*

J. Schwartz[#], F. Hunte, W.K. Chan, X.F. Gou, X.T. Liu, M. Phillips, Q.V. Le, G. Naderi, M. Turenne, L. Ye, Department of Materials Science and Engineering, North Carolina State University, Raleigh, NC 27695, U.S.A

*Abstract*

This paper reviews the status of high temperature superconductors for high field magnets for future devices such as a high energy LHC or a muon collider. Some of the primary challenges faced for the implementation of systems are discussed. Two conductor technologies, $Bi_2Sr_2CaCu_2O_{8+x}$ and $YBa_2Cu_3O_{7-\delta}$, have emerged as high field conductor options, but their relative advantages and disadvantages for high field magnets are quite different. These are reviewed from an engineering perspective, including coil manufacturing, electromechanical behaviour and quench behaviour. Lastly, the important roles of "system pull" upon conductor and magnet technology development, and of interactions between the materials and magnet communities for accelerating development, are discussed.

## INTRODUCTION

High temperature superconductors (HTS) have continued to advance technologically such that there are now at least six demonstrations of the generation of magnetic field greater than 25 T, and at least two that have surpassed 30 T [1, 2]. With the successful operation of the LHC, it is timely to consider the technological prospects for the development of the large, high field superconducting magnets needed for the next generation of colliders, such as a high-energy upgrade to the LHC or a muon collider [3, 4]. As HTS conductors evolve into commercial products, it is also important to assess the technological limitations and challenges that need to be addressed for large systems to ultimately come to fruition. Furthermore, with the high cost associated with development of future magnet technologies, an assessment of decision-points is also appropriate.

## CONDUCTOR OPTIONS

As discussed at length in [5], the use of HTS conductors for high field magnets is necessitated by the fundamental limit to the high field behaviour of $Nb_3Sn$. Thus, although $Nb_3Sn$ currently has significant advantages over HTS conductors in terms of cost, availability, experience-base and fundamental understanding, it is limited to magnets generating about 21 T for solenoids, and perhaps 18 T for dipoles. It is only the ability to carry high critical current density ($J_c$) at very high magnetic field (at least 45 T) that results in the consideration of HTS options. Thus, HTS conductors are viewed not as a replacement technology, but as an enabling technology for future high field magnet systems. Figure 1 plots the engineering critical current density versus magnetic field for LTS, HTS and $MgB_2$ conductors [6]. This data represents the highest published values for each of the emerging conductor options. Note that the high field performance of $MgB_2$ is poor, so it is not considered a high field conductor option.

### Emerging Conductor: $REBa_2Cu_3O_{7-\delta}$

$REBa_2Cu_3O_{7-\delta}$, (REBCO), where RE refers to rare earth elements, is an HTS conductor that has been developed via thin film oxide technologies. While there are variations from manufacturer to manufacturer, in general REBCO conductors are based upon the deposition of thin oxide buffer layers atop a high strength Ni-alloy (e.g., Hastelloy or Ni-W) substrate. The REBCO layer is then deposited upon the oxide buffer layers, which provide a template for bi-axially-textured growth and a chemical barrier against Ni contamination. The bi-axial texture is known to be essential for obtaining high $J_c$. The REBCO layer is then covered by a thin Ag "cap layer" that provides environmental protection. Lastly, the entire conductor is encased by stabilizer, typically Cu. The resulting "coated conductor" carries the highest high-field $J_c$ of any known superconducting material. The REBCO fill factor is only ~1-2%, however, which greatly reduces the engineering critical current density, $J_e$. Extensive literature exists regarding the processing of REBCO conductors, and various approaches used to enhance flux pinning, mechanical strength, REBCO layer thickness, etc. [7-14]

One of the primary limitations of REBCO coated conductors is that excellent electromagnetic performance is only obtained in a highly aspected, wide, thin tape geometry with highly anisotropic electromagnetic behaviour. The anisotropy limitations can be overcome to a significant extent in solenoids by using the REBCO only in the highest field section of the magnet system, and using NbTi and $Nb_3Sn$ outserts to generate the lower magnetic fields. By properly designing the relative heights of the outserts and the REBCO insert, the magnetic field perpendicular to the REBCO tape (the "bad" direction) can be minimized and overall magnet performance optimized [15]. This is not so readily accomplished for dipoles or quadrupoles magnets.

Another challenge for REBCO is that the wide, thin tape geometry does not readily lend itself to traditional cabling, and Rutherford cables are not an option at present. The only cabling option currently available is the Roebel cable approach, which is intrinsically more expensive as it wastes a significant percentage of the REBCO conductor in its manufacture. Although Roebel cables are clearly a low AC loss option, they are unproven for high field magnet systems, and in particular their mechanical and quench behaviours are unknown [16-19].

___________________________________________
*Work partially supported by the U.S. Department of Energy through the Very High Field Superconducting Magnet Collaboration, the SBIR program and the STTR program, in collaboration with Supercon Inc. and Muons Inc.
[#]justin_schwartz@ncsu.edu

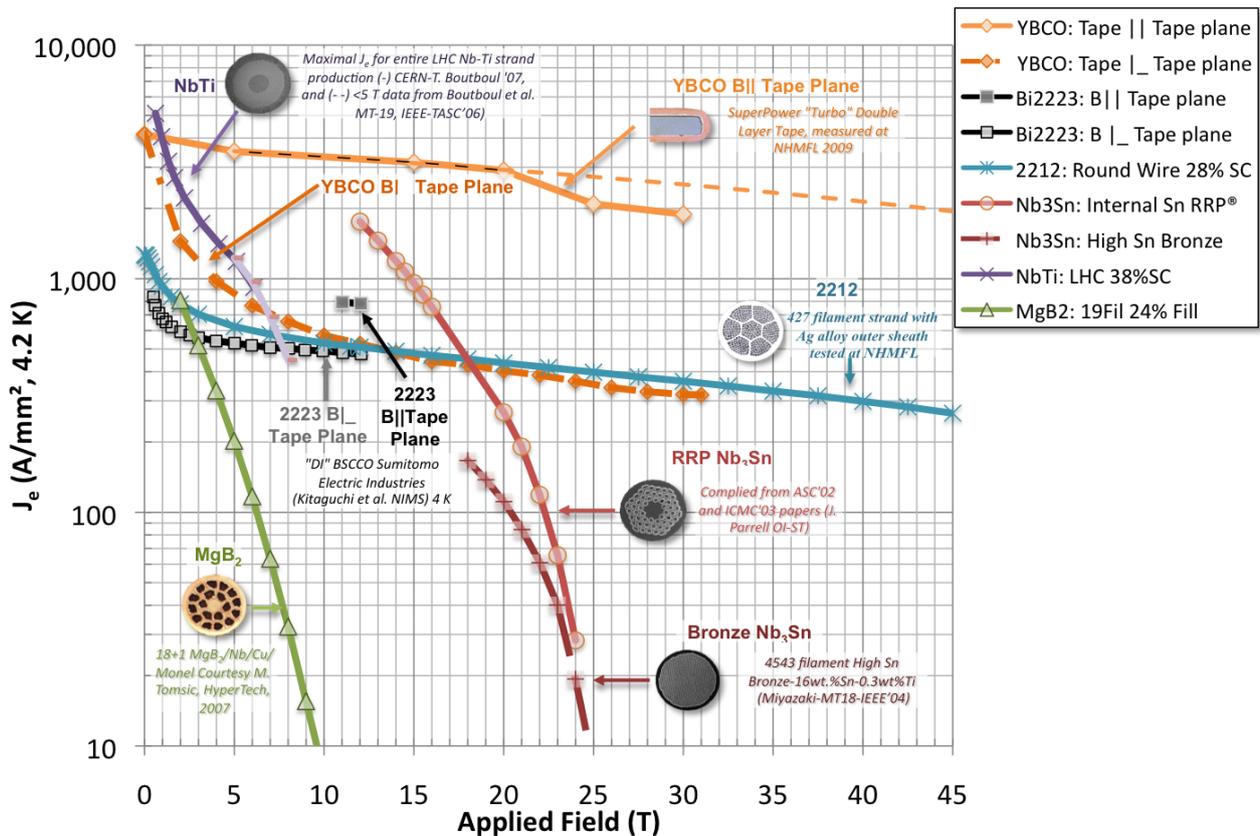

Figure 1: Engineering critical current density versus magnetic field for low temperature superconductor wires, high temperature superconductor wires and tapes and $MgB_2$ wires.

REBCO coated conductors can be viewed as a commercial product, although production lengths remain limited and at present the cost is high. Industrial manufacturers are currently focusing significant attention on scale-up issues in anticipation of meeting demand for a number of potentially growing markets, particularly within the energy sector. As the expectation is that the high volume market will ultimately be for energy systems operating at temperatures approaching 77 K and at relatively low magnetic field, optimizing conductors for low temperature, high field operation is not a research and development (R&D) priority. Thus, while the anticipation of market pull is driving development and scale-up, the impact on high field magnets is not as great as if there was a dedicated focus specifically upon the development of high field conductors. The potential for a large commercial market, coupled with the relatively low raw materials costs, imply that as commercialization increases and the REBCO market grows, the conductor unit price will decrease significantly. In the short term, REBCO applications will be attractive when REBCO is the enabling technology, because it is not cost competitive with LTS materials, but if sufficient demand is established such that the unit price is reduced significantly, then that price reduction is likely to fuel additional demand and a growing market.

### Emerging Conductor: $Bi_2Sr_2CaCu_2O_{8+x}$

Unlike REBCO, $Bi_2Sr_2CaCu_2O_{8+x}$ (Bi2212) conductors are based on powder-in-tube technology and multifilamentary wire deformation processes developed for NbTi and $Nb_3Sn$. Both single and double restack architectures are manufactured industrially, but in most cases the first billet uses a high purity Ag tube, and the subsequent tubes are a Ag-alloy, typically Ag-Mg. The starting powder is typically high phase purity Bi2212. After deformation, the multifilamentary wire requires a heat treatment that first goes above the peritectic melt temperature and then resolidifies the Bi2212 phase. The partial-melt process is necessary to establish connectivity between the Bi2212 grains. During peritectic melting, however, phase segregation occurs so after resolidification the oxide is not phase-pure Bi2212, but instead contains a number of parasitic phases that reduce wire performance. Furthermore, the powder-in-tube process does not result in 100% dense filaments, so after heat treatment there is also significant porosity within the filaments. Recent studies have shown that the heat treatment results in interfilamentary bridging which plays an important role in transport as well. Extensive literature exists regarding the processing of Bi2212 conductors [20-32].

Despite these issues, Bi2212 round wire remains a strong candidate for high field magnets. Bi2212 wires are electromagnetically isotropic and readily formed into Rutherford cables [27, 33, 34]. Industrial wire production is primarily limited by demand; the manufacturers are capable of producing significant volumes upon order. The key supply-chain concern, however, is the Bi2212 powder itself, but this is also a demand-based issue. If significant volumes of Bi2212 are required, capable powder manufacturing exists. The lack of significant demand for Bi2212 is primarily a result of the lack of a market other than high field magnets. The in-field electrical performance of Bi2212 declines rapidly for temperatures above about 20 K, so they are not competitive with REBCO for applications within the energy sector. Thus, while Bi2212 R&D is focused on high field magnets, there is no other strong driving force for scale-up. Unlike REBCO coated conductors, the unit costs of Bi2212 are not dominated by the manufacturing costs but instead by the unit cost of Ag. Thus, the price of Ag represents the "floor" below which the price of Bi2212 wire cannot drop, and the only potential for decreasing the conductor cost for Bi2212 magnets is to significantly increase $J_e$ such that less conductor is required.

*Conductor Comparison*

Although REBCO and Bi2212 are both HTS conductors, technologically their similarities are few and their R&D challenges for high field magnets are quite different. This is summarized in Table 1, which compares the two conductors in terms of magnet-related issues.

## ELECTROMECHANICAL BEHAVIOUR

High field magnets are intrinsically high force magnets due to the Lorentz forces present. Thus, strain tolerance and strain management grow in importance as the magnetic field increases. While low current density is one approach to lower Lorentz forces, low current density also results in very large, expensive magnets. Thus, an ideal high-field conductor not only has high $J_c(B)$, but also either strain tolerance or compatibility with approaches to reduce the conductor strain in magnets.

*Electromechanical behaviour of REBCO*

REBCO is manufactured on strong Ni-alloy substrates that provide significant mechanical advantages. With one of the approaches to REBCO, Hastelloy is used as the substrate and the Cu-stabilizer is attached to the conductor via electroplating. These approaches result in a particularly robust conductor and mechanical limitations are not a primary concern. The only uncertainty is in regards to tensile loads normal to the wide face of the conductor. The alternative approach to REBCO conductors is a Ni-W substrate with stabilizers attached via solder fillets. While Ni-W is stronger than most other high-field conductor options, it is not as strong as Hastelloy, and the solder fillets do not provide high strength for tensile loads normal to the wide face of the conductor or in shear. Thus, of the two conductor technologies, the former is preferred for high field magnets [35-42].

Table 1: HTS Conductor Comparison
(Note that bold text indicates a significant advantage whereas italics indicates a particularly challenging issue)

| Bi2212 | REBCO |
|---|---|
| **Round wire** | *Wide, thin tape* |
| ~30% fill factor | *~1-2% fill factor* |
| **Isotropic electromagnetic behaviour** | *Anisotropic electromagnetic behaviour* |
| **Rutherford cables** | *Roebel cable only* |
| *Weak Ag-alloy matrix* | **Strong Ni-alloy substrate** |
| Poorly understood microstructure-property relationships; properties very sensitive to heat treatment details | Highly engineered microstructures |
| **Readily scalable conductor manufacturing** | *Conductor manufacturing scale-up challenges* |
| *Wind&react magnets; magnet processing challenges* | **React&wind magnets** |
| High field magnet applications are sole market | High temperature, low-field applications driving R&D |
| *High price of Ag* | *Expensive processing* |
| **Active high field magnet projects on-going** | **Active high field magnet projects on-going** |

REBCO magnets are limited to react-and-wind magnet manufacturing which can add a non-zero bending strain to the total strain. The conductor can be designed, however, such that the REBCO layer is situated on the neutral axis or even in compression, so the bending strain contribution to the total strain is not a dominant concern [43]. Furthermore, REBCO electromechanical behaviour is reversible, so cycling within the fatigue limits does not generally result in significant degradation before failure [44, 45].

*Electromechanical behaviour of Bi2212*

Unlike REBCO, Bi2212 is encased within a relatively weak Ag/Ag-alloy sheath that provides significant ductility for wire drawing but does not provide high strength. Typically, the Ag-alloy is oxide dispersion strengthened (ODS) Ag-Mg, which sacrifices some ductility for increased strength and stiffness after heat treatment, but Bi2212 wire electromechanical behaviour remains a serious drawback. Primarily due to the poor electromechanical behaviour, Bi2212 is limited to wind-and-react magnet construction, which eliminates bending strain but results in other limitations discussed below [46-48].

While the primary source of the poor electromechanical behaviour of Bi2212 is the lack of strength in the Ag/AgX matrix, the post-heat treatment Bi2212 microstructure is also a significantly contributing factor. Reacted Bi2212 multifilamentary wire microstructures are comprised of discontinuous filaments, interfilamentary bridges, non-superconducting oxide phases and porosity. As a result, the microstructure is defect-intensive with a ready supply of potential crack initiation sites that can lead to poor strain tolerance. This has recently been confirmed by statistical analysis that shows significant variance (i.e., inhomogeneity) in performance [49, 50]. This is illustrated in Figure 2, which shows Weibull reliability distribution for Bi2212 round wires tested at three different strain values (zero strain, 0.25% and 0.40%). While the high current "tail" in the 0.40% strain data indicates a high-strength electrical network within Bi2212, the reduction in reliability of the 0.25% strain relative to the zero-strain is indicative of some degree of irreversibility. The same analysis on REBCO coated conductors did not show similar inhomogeneous behaviour. Furthermore, there is evidence that the Bi2212 filaments show fractal characteristics, and subsequent fractal analysis highlights the role of these defects in the electromechanical performance.

There are significant R&D efforts currently aimed at improving Bi2212 performance. One research direction is aimed at improving the mechanical properties of the AgX sheath. By increasing the sheath stiffness, the strain on the superconductor is reduced for a constant load. Another research direction is aimed at improving Jc through improved processing and heat treatment approaches that result in an improved microstructure. If the inhomogeneous Bi2212 microstructure is a limiting factor in both Jc and the electromechanical behaviour, then it is anticipated that as Jc increases, the strain tolerance will also improve. There is recent evidence that Bi2223 (and thus perhaps Bi2212) has an underlying reversible component to its electromechanical behaviour, which may indicate the potential for significant improvements [51].

## COIL MANUFACTURING

As indicated in Table 1, REBCO magnets are wound using the react-and-wind approach, simplifying the selection of insulation and instrumentation materials. Although some of the allowed strain may need to be allocated to bending strain, this has minimal impact on magnet design because the Ni-alloy substrate has high strength, and because the REBCO layer is on or near the conductor neutral axis. If the REBCO layer becomes substantially thicker, if double-sided coating of the Ni-alloy substrate becomes a standard technique for increasing $J_e$, or if a multilayer approach is developed, then the bending strain could become significant. At present, however, magnet manufacturing is not one of the primary challenges to high field REBCO magnets.

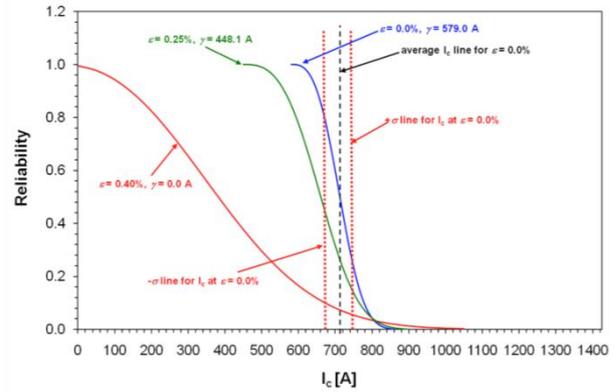

Figure 2: Weibull reliability distribution functions for Bi2212 round wire. The red curve is for 0.40% strain, the green is for 0.25% strain and the blue is for zero strain. For more details, see [49, 50].

Primarily due to their strain sensitivity, Bi2212 magnets are limited to wind-and-react (W&R) manufacturing or variations thereof [22, 23, 52, 53]. W&R manufacturing is not intrinsically problematic, as there is a large experience base to draw upon from Nb$_3$Sn magnets and a large number of very large magnet systems have been manufactured in this manner. W&R manufacturing of Bi2212, however, is significantly more challenging than Nb$_3$Sn for two unavoidable reasons: the presence of oxygen, which limits the options for insulation and instrumentation materials, and the high degree of sensitivity of the electrical performance of Bi2212 to the details of the heat treatment. In particular, due to the peritectic melting/resolidificaiton process required for high $J_c$ Bi2212, $J_c$ is very sensitive to a narrow temperature window (2-3 $^{\circ}$C) for the peak heat treatment temperature and the amount of time spent above the peritectic melt temperature. The latter challenge is exacerbated as the size of coils increases. It is necessary to ensure that the portion of the magnet that is slowest to reach the peak temperature experiences the peritectic melting; but while waiting for the thermal diffusion, the portion of the magnet that is first to experience melting remains above the melt temperature for too long.

As a result of these challenges, insulated short samples often show a reduced $I_c$ relative to bare wires (about 10% lower), and even relatively small Bi2212 magnets consistently show about 30% lower $I_c$ than bare short samples (see Figure 3). Some of this decrease is likely due to the presence of insulation that can enhance the depletion of Cu from the conductor (without insulation, some Cu diffuses from the filaments into the Ag-alloy; reactions with the insulation provide an additional sink for Cu that diffuses to the edge of the wire).

Solutions to these challenges are currently under development within the Bi2212 community. Alternative heat treatment processes that reduce the sensitivity upon the heat treatment peak temperature are being investigated, as are alternatives that attempt to minimize the presence of porosity and parasitic oxide phases. Alternative sheath alloys that deter Cu diffusion from the oxide cores, and alternative insulation materials that are not Cu-getters and are significantly thinner than presently available options, are being developed. Lastly, new heat treatment monitors that provide a continuous temperature-time map within the magnet during heat treatment offer the prospect of knowing real-time when the coldest portion of the magnet reaches the peritectic melt, facilitating dynamic control of the magnet heat treatment process such that the time above the resolidification temperature can be actively engineered [54, 55].

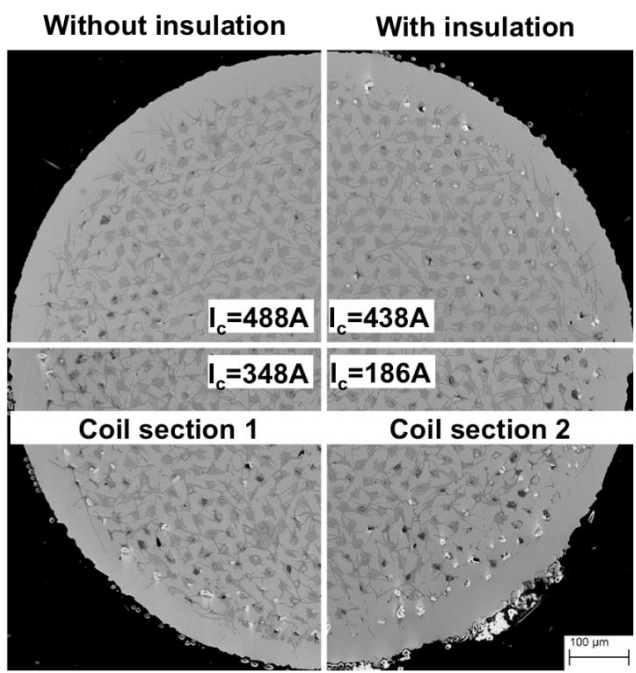

Figure 3: SEM images of heat treated Bi2212 round wire sections. The upper two images are from short samples (without and with insulation) and the lower two are from sections of an insulated Bi2212 coil. The corresponding critical currents are also shown. All samples were heat treated together during the same furnace run.

## QUENCH BEHAVIOUR

An important issue for any large, high-energy superconducting magnet system is quench protection. For low-temperature superconductor (LTS) based magnets, quench protection is well understood and protection techniques are well established. For HTS-based magnets, the underlying science is qualitatively similar to that of LTS magnets, but the behaviour is quantitatively very different and thus new approaches may be required.

Typically, quench protection involves (1) quench detection: identifying that the magnet is going to quench while rejecting false signals and disturbances from which the magnet can recover, and (2) a protective response that must be implemented on a time scale fast enough to prevent the magnet from being damaged. Thus, to design a quench protection system, the magnet designer must needs to consider the time-evolution of the voltages and temperatures within the magnet during a quench so effective quench detection schemes can be designed, and the voltage-driven and temperature-driven degradation limits so that the protective response can prevent damage to the magnet. Note also that as the magnet stored energy increases, so does the risk of degradation during a quench. As the stored energy is proportional to $B^2$, the importance of quench protection increases with magnetic field and with magnet cost.

One of the most important differences between LTS and HTS magnets is the quench propagation velocity (QPV), which is a key parameter for quench detection. It has been consistently shown that the QPV in HTS (Bi2212 and REBCO) is significantly slower, as much as two orders of magnitude slower, than in $Nb_3Sn$ magnets (see Figure 4) [56-69]. While slow quench propagation may also be correlated with a slower local temperature rise within the magnet, the key question is related to the rate of *localized* temperature rise as compared to the rate of voltage rise over monitored segments of the magnet. Voltage is, by definition, the integral of the electric field over the monitored length of conductor. It does not consider the spatial profile of the electric field (which is directly related to the spatial profile of the temperature in the conductor). Thus, since slow propagation results in a highly peaked temperature profile, for the same voltage, the peak temperature in an HTS magnet is likely to be much higher than in an LTS magnet. As a result, one of the key challenges for large, high field, HTS magnets is quench detection.

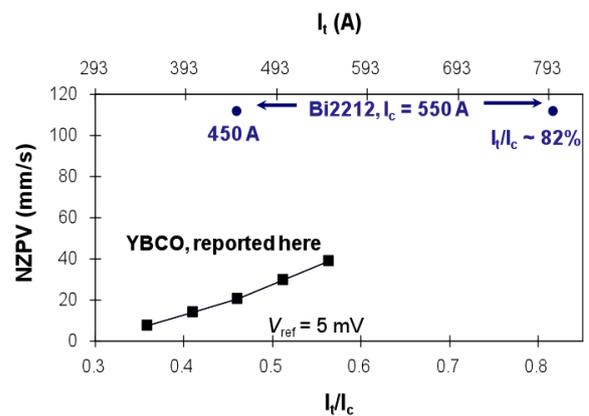

Figure 4: Experimentally measured normal zone propagation velocities of Bi2212 and YBCO coated conductors (from [60]).

A number of approaches are under development to address the HTS quench detection challenge. One approach is the development of quench detection sensors that are not voltage-based. The leading option in this regard is the implementation of optical fibers [54, 55, 70-

77]. Optical fiber sensors are in-service in a number of environments today, but as of yet have not been optimized for low temperatures. Optical fibers are small, thin fibers that can be directly incorporated into an HTS winding. Compatibility with the Bi2212 heat treatment has been demonstrated from the perspective of not causing any degradation to the Bi2212 wire. There are a number of approaches that use optical fibers as sensors, including fiber Bragg gratings, Brouillion scattering, and Rayleigh scattering. Fiber Bragg gratings are the most commonplace and the most developed of the optical fiber sensor technologies, however they are point sensors. While one fiber can be used to make measurements at a large number of points along its length, the locations must be predetermined so that gratings can be written into the fiber. Furthermore, while optical fibers in general survive the Bi2212 heat treatment, there are problems with survival of the grating itself. Rayleigh scattering is similar to fiber Bragg grating based scattering, but rather than using an engineered grating, it relies upon the natural inhomogeneities within an individual fiber to provide the necessary light scattering. Thus, in principle, the limit to the spatial resolution in Rayleigh scattering is the wavelength of the light source used. As a result, the practical resolution limits (spatial and temporal) are related to the data acquisition and data analysis. Quench detection requires obtaining a series of scattering profiles from the length of fiber and comparing them to determine if there are significant changes occurring. To effectively implement such a system, the required spatial and temporal resolutions must be understood and the data acquisition and analysis developed to meet those requirements.

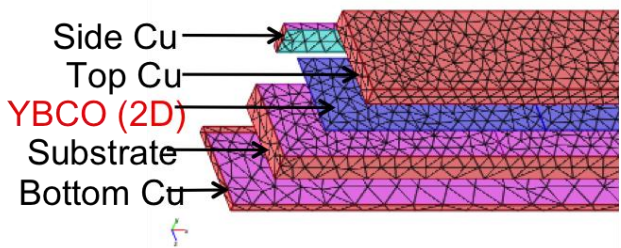

Figure 5: Meshing of a YBCO coated conductor.

In order to better understand the spatial and temporal resolution requirements for quench detection, and the microscopic behaviours during a quench (localized temperatures, voltages, current distribution, stresses and strains), and the likely failure modes during a quench, and thus to assist magnet designers in developing effective quench protection schemes, a high fidelity, experimentally validated, multiscale model of REBCO coated conductor quench behaviour has been developed. A typical meshing of a REBCO conductor is shown in Figure 5, and a resulting plot of the temperature versus location during a quench is shown in Figure 6. This important tool, which is already providing insight into the engineering of REBCO architectures for improved quench performance, is also capable of predicting both macroscopic, three-dimensional quench propagation within a magnet while simultaneously monitoring the localized behaviour within the microscopic layers of a REBCO conductor. A mesh of a coil section, with an embedded microscopic mesh of a section of conductor, is shown in Figure 6 [59, 61].

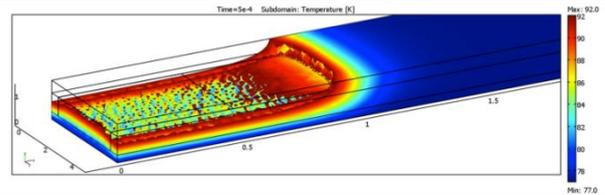

Figure 6: Temperature versus location during a quenching YBCO conductor.

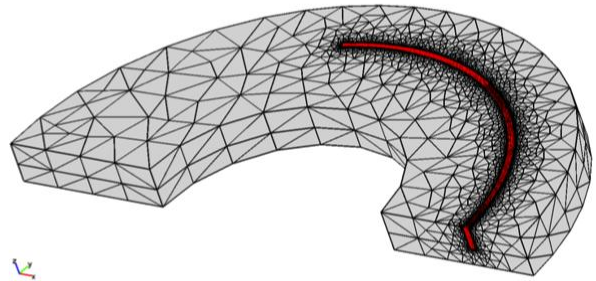

Figure 7: Multiscale model showing a course mesh for a section of a coil with a finer-scale mesh (micron-scale) for a section of YBCO conductor within the coil.

One of the key results from the three-dimensional modeling is that three-dimensional quench propagation within a magnet can significantly reduce the peak temperature for a fixed voltage within a coil. This may be obtained via the development of thermally-conducting electrical insulators to serve as turn-to-turn insulation. While it is found that the one-dimensional propagation velocity (along the conductor length) is decreased, the peak temperature in the magnet is also significantly reduced. This is illustrated in Figure 8, which shows the peak temperature (fixed voltaged) for three different insulation options. The blue (highest peak temperature) is for kapton, the black (intermediate temperature) is for alumina, and the red (lowest temperature) is for a high thermal conductivity alternative [78].

The other key question for quench protection is understanding the failure limits. For REBCO conductors, the primary concern is delamination. For conductors manufactured with solder fillets, the melting temperature of the solder is the primary concern. For electroplated conductors, two degradation mechanisms have been identified. The first is related to pre-existing defects in the conductor. These are thus a manufacturing issue that can be alleviated with improved processing, quality analysis and quality control. The second is related to delamination at the REBCO/Ag interface and has been observed during quenching [79]. Ultimately this is likely to be the fundamental limit. If this limitation can be addressed, then the most fundamental limit becomes deoxygenation of the REBCO itself. For Bi2212, the quench limits are directly related to the electromechanical limits; i.e., the local stress and strain within the conductor [64]. As with the electromechanical behaviour, there is anticipation that with with improved Bi2212 microstructure and stiffer AgX sheaths, improved resistance to quench-induced degradation will also result.

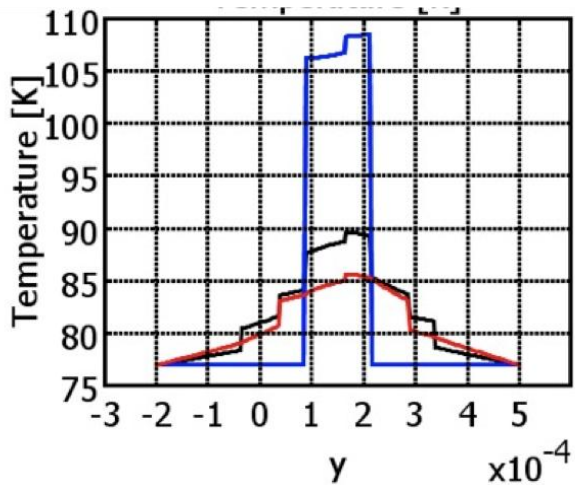

Figure 8: Impact of thermal conductivity of electrical insulator on the peak temperature within a YBCO coil during a quench.

## OTHER CONSIDERATIONS

Although conductor options, electromechanical behaviour, coil manufacturing and quench behaviour are the primary issues for developing future high-field HTS magnets, other issues can also play important roles in the development and implementation of new conductor technology for accelerator magnets. For example, large magnets will require long lengths of conductor, as well as connections to current leads. Thus, joining technologies are important. These have not been effectively developed for either Bi2212 or REBCO conductors. Furthermore, the effects of irradiation on Bi2212 and REBCO magnets are not yet known. From a quench perspective, HTS magnets tend to be very stable with large minimum quench energies, so they may be ideally suited for regions of the system with large irradiation heat loads. But this is predicated upon resistance to irradiation damage, both for the conductor and any other materials (e.g. insulation) within the magnet system. Furthermore, for magnets to be used in high irradiation areas (e.g., an interaction region), HTS magnets, and in particular REBCO magnets, may offer the option of having the operating temperature as a design variable. In general, at low temperature the energy margin increases with temperature, so there may be situations where a higher operating temperature is preferred because it results in a larger temperature margin, even at the expense of some critical current density.

Another important consideration in the development of high field magnets is the impact of "magnet pull". Progress in the development of HTS technologies specifically aimed at the low temperature, high field regime is likely to be directly correlated with demand for such technologies by the magnet communities. In the over 20 years of HTS conductor development, most of the progress has been via "conductor push", with progress in the materials science of HTS conductors coming without consistent guidance regarding the specific demands of real magnet applications. In recent years, however, magnet pull has become an increasing presence in the development of HTS technology, and such pull is likely to continue to have significant impact on progress.

Lastly, it is important to consider potential "game changers" that could transform high field magnet technology. For example, if REBCO could be manufactured as an *isotropic round wire* with the same electrical and electromechanical performance as present-day REBCO coated conductors, Bi2212 would probably be eliminated from consideration and the primary R&D focus would concentrate REBCO scale-up, quench detection, joining, etc. Similarly, if Bi2212 could be manufactured with 100% dense, continuous, phase-pure filaments in a high strength, high stiffness sheath, then Bi2212 would likely leapfrog ahead of REBCO conductors and a new high field magnet technology would emerge.

## SUMMARY AND CONCLUSIONS

Future devices for high-energy physics are likely to require magnetic fields greater than that which $Nb_3Sn$ technology is capable of generating. As a result, new magnet technologies based upon HTS materials, primarily REBCO and Bi2212, are likely to be needed. These conductors are progressing and demonstrations of magnetic field generation greater than 25 T have been achieved repeatedly. Both conductor technologies, however, have significant remaining hurdles that must be overcome before the next generation of devices can be constructed. Although in many ways Bi2212 and REBCO are similar, each has distinctly different strengths and weaknesses, and thus the R&D programs required for each are quite different. For Bi2212, the primary challenges are the electromechanical behavior and large magnet manufacturing and heat treatment. For REBCO, the primary conductor challenges are the very low fill factor, electromagnetic and geometric anisotropy, and scale-up. Furthermore, both conductors show quench behaviour that is quantitatively quite different from that of LTS magnets.

Despite these challenges, Bi2212 and REBCO have made significant progress in recent years due to the presence of "magnet pull". Low temperature, high field magnets are needed for future nuclear magnetic resonance devices, future high energy physics devices, and recently even future energy storage devices. The interest from these communities is having an important effect on the development of conductor and magnet technologies focused on the low temperature, high field regime.

round wire," *IEEE Transactions on Applied Superconductivity,* vol. 19, pp. 3061-3066, 2009.